\documentclass[aps,prl,twocolumn,showpacs,preprintnumbers,amsmath,amssymb,superscriptaddress]{revtex4-1}
\usepackage{graphicx}
\usepackage{epsfig,bm}
\usepackage{color}

\usepackage{amssymb}
\usepackage{mathrsfs}
\usepackage{amsmath}
\usepackage{subfig}
\usepackage{hyperref}
\usepackage[normalem]{ulem}

\begin{document}

\title{Passive viscoelastic response of striated muscles}

\author{Fabio Staniscia}
\email{fabiostaniscia@gmail.com}
\affiliation{LMS, \'Ecole polytechnique, 91128 Palaiseau Cedex, France}
\author{Lev Truskinovsky}
\affiliation{PMMH, CNRS - UMR 7636 PSL-ESPCI, 10 Rue Vauquelin, 75005 Paris, France}


\date{\today}
\begin{abstract}	
Muscle cells with sarcomeric structure exhibit highly nontrivial passive mechanical response.  The difficulty  of its continuum modeling is due to the  presence of long-range interactions transmitted by extended protein skeleton. To build a rheological model for muscle 'material' we use a stochastic micromodel and derive  a linear response theory for a  half-sarcomere.  Instead of the first order  rheological equation, anticipated by  A.V. Hill on the  phenomenological grounds,  we obtain a novel second order  equation.  We use the  values of the microscopic parameters for frog muscles to show that the proposed rheological model is in excellent quantitative agreement with physiological experiments.  
\end{abstract}
\maketitle

One of the simplest  biological systems,  that still defies the  attempts to   reproduce it  artificially  as a  macroscopic material,  is a striated muscle~\cite{Narayanan2018}.   Its mechanical complexity  is due to the presence of a large number of nonlinear,  hierarchically organized microscopic sub-systems  that are  strongly coupled  through long-range interactions~\cite{Caruel_2018}. This makes the task of  reconstructing the macroscopic   constitutive  relations  describing even its passive mechanical response rather challenging~\cite{Caruel2019,Regazzoni2018}.

A broadly used  phenomenological theory of the passive  visco-elastic response  of striated muscles,  proposed by A.V.Hill~\cite{Hill38,hill1949,GLANTZ1974137,Gerazov16},  does  not rely on  coarse graining techniques~\cite{DeVault65,Bavaud1987} and therefore does not offer a  link between macro and micro parameters. Since Hill's rheological relation  involves a single characteristic time scale,  it  also does not capture the  difference in the   passive response exhibited by striated muscles abruptly loaded in soft (isotonic) and hard (isometric) loading devices~\cite{Huxley71, Reconditi04}.

A microscopically guided stochastic approach to muscle visco-elasticity was proposed by Huxley and Simmons~\cite{Huxley71,Eisenberg80} who assumed that the individual force producing units (myosin cross-bridges) are stochastically independent.  A mean-field interaction between the cross-bridges was incorporated in  a closely related  model  by  Shimizu~\cite{Shimizu72,Kometani1975,Bonilla87,Frank04}. The two approaches have been  recently unified~\cite{Marcucci2010,Marcucci2010b,Caruel13,Caruel_2018}. In  the present article we use this framework to rigorously  derive  from a micro-model  a linear rheological  response theory for a muscle half-sarcomere. Our analysis builds on the work of Shiino~\cite{Shiino87} who obtained  a similar linear response theory for the related  model of Desai and Zwanzig~\cite{Desai1978,Bonilla87,Frank04};  other relevant out-of-equilibrium systems were studied in~\cite{Patelli2012,Patelli2014}.
  
Our  main result is the   linear  spring-dashpot   scheme  which reproduces  adequately the mechanical behavior  of a muscle fiber subjected to a time dependent perturbation. In contrast to the classical model of Hill \cite{Hill38,hill1949}, the proposed rheological equation  contains not only the \emph{first} but also the \emph{second } time derivatives of the macroscopic displacement.  We use the  values of the microscopic parameters for frog muscles to show that our  macroscopic model,  which does not rely on any  fitting parameters, is in excellent quantitative agreement with physiological experiment.  

\begin{figure}[h]
\includegraphics[width=0.5\textwidth]{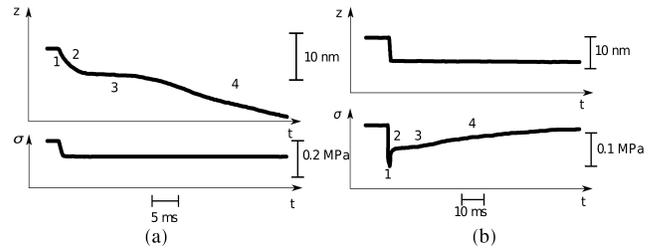}
\caption{Schematic representation of the transient response of a skeletal muscle subjected to an abrupt (a) isotonic and  (b)  isometric  perturbation.   Here $\sigma$ is the stress  and $z$ is the elongation. The meaning of phases 1-4 is explained in the text, and the focus of this article is on phases 1 and 2. Adapted from \cite{Huxley74}.} \label{fig:exp_schem}
\end{figure}

The model contains two time scales that can be  associated with the transient stages of  muscle response known as phases 1 and 2, see  Fig.~\ref{fig:exp_schem} and \cite{Huxley71,Piazzesi03,Decostre05,Piazzesi07},   with both of them   captured   adequately by our rheological model.    The presence of the two time scales  reflects  involvement   of the two parallel \emph{passive}  processes: the (microscopic) conformational change in $N$ myosin heads and the macroscopic  relaxation of the myofilaments in a viscous environment.  The obtained   linear rheological equation does not capture the collective barrier crossing  which can be associated with the slower phase 3 explained in~\cite{Caruel16,Caruel17,Caruel_2018};  it also does not address the \emph{active} phase 4, see Fig.~\ref{fig:exp_schem}.

We recall that striated muscle is  a hierarchical chemo-mechanical system  with the  smallest scale represented by force generating half-sarcomeres~\cite{AlbertsBook13,Caruel_2018}. To reproduce  the passive response of  this molecular machine  we use  the simplest  microscopic model developed in~\cite{Marcucci2010,Marcucci2010b,Caruel13,Caruel16,Caruel17}. 
We assume that behind the  muscle  power stroke, responsible for fast force recovery, is  a   double well potential of Landau type, $V(x)$, describing two conformational states of  a cross-bridge~\cite{Marcucci2010}; it is known that   myosin heads can be in other  configurations~\cite{Huxley74,Decostre05,Caremani13}, so here we simplify the real physical picture \cite{Molloy1995,Reconditi04}. We also assume  that  the potential $V(x)$ is asymmetric, which allows the system to generate (stall) force in the  physiological regime of isometric contractions.   While this asymmetry is maintained  actively  \cite{Sheshka16},  we can still interpret the  short time response in such system as passive. The active response involving detachment of the cross-bridges becomes dominant  at time-scales of the order of $40$ ms~\cite{howard2001}.

 To define the  dimensionless units we  rescale the length by the  size of the maximum  working stroke $a$, the energy by $\kappa \, a^2$, where $\kappa$ is the cross-bridge stiffness, and the time  by the ratio of  the cross-bridge drag coefficient   $\gamma_x$ and $\kappa$.
To model a bundle of thick and thin filaments  linked by  $N$ cross-bridges and loaded with the force  $f$  we use the dimensionless energy~\cite{Caruel_2018}   
\begin{equation}
\label{eq:ener_hard_dev}
    H=\sum_{i = 1}^N \left[ V(x_i) + \frac{ (y - x_i)^2 }{2} +  \frac{\lambda_f}{2}  \left(z - y \right)^2\right] - z\,f, 
\end{equation}
 where the  variables $x_i$ represent the configuration of  individual cross bridges.  The latter are elastically coupled through the cross-bridge stiffness  to a collective variable $y$ representing  an actin filament; the associated quadratic term in \eqref{eq:ener_hard_dev}  represents mean field type interaction.  The combined elasticity of actin and myosin filaments is described by the quadratic term coupling the variable $y$ and $z$ with the  dimensionless coefficient $\lambda_f$ representing the overall filamental stiffness, see  \cite{Caruel_2018}  for more details.

We assume that the meso-scopic collective variable $y(t)$ relaxes instantaneously~\cite{HUXLEY19942411,WAKABAYASHI19942422} and we can therefore  eliminate such  (Weiss-type)  variable  adiabatically. We obtain $y =    (\langle x\rangle + \lambda_f \, z)/(1 +\lambda_f),$  
 where we introduced notation $\langle x\rangle= N^{-1} \, \sum_{i=1}^N x_i .$  Substituting the obtained expression for $y$  into \eqref{eq:ener_hard_dev} we obtain the redressed energy $\tilde H(\textbf{x},z)=H (\textbf{x} ,y( \textbf{x},z),z)$ whose  main feature vis-\`a-vis the model of Huxley and Simmons   is the presence of  all-to-all interaction between the cross-bridges  expressed through the term  $ \langle x\rangle^2$, see also \cite{Desai1978,Shiino87}.

Next we  assume that the macro-variable $z(t)$ is also deterministic, however now its dynamics is governed by the relaxation equation:
$
  \nu  \dot{z} = -    \partial   \tilde H/\partial   z,  
   $ where the superimposed dot denotes time derivative and $\nu$ is the  dimensionless  macroscopic friction coefficient  obtained by normalizing the filamental drag coefficient $\gamma_z$  by the myosin head drag coefficient $\gamma_x$.   We can rewrite this equation as
\begin{equation}\label{eq:sig_vs_z_dyn}
\frac{\nu}{N } \, \dot z  =     \lambda_f \frac{\langle x\rangle -  z }{1+ \lambda_f}  + \frac{  f }{ N} .
\end{equation} 
The time scale  $\tau_z = \nu  (1+\lambda_f)/(\lambda_f  N)$ will then characterize the  evolution of the macro-variable $z$.  
 
In the evolution of the micro-variables  $x_i$ we need to account for the thermal noise:  
$
 \dot{x}_i = -  \partial \, \tilde H/\partial \, x_i + \xi_i , 
$
where  $\langle \xi_i \rangle =0$ and  $\langle \xi_i(t) \, \xi_j(t^{\prime}) \rangle = (2/\beta) \, \delta(t - t^{\prime}) \, \delta_{i j} $. Here we introduced  the inverse dimensionless temperature  $\beta$.  Under the assumption that $N$ is large, the single particle probability density $p(x,t)$  can be found from the nonlinear Fokker-Planck equation~\cite{Desai1978,Shiino87,FrankBook05}
\begin{multline}
 \label{eq:fokker-planck}
 \frac{\partial p }{\partial t} =  \frac{\partial}{\partial x} \left[ \left(  \frac{\partial V}{\partial x}  + x - \frac{\langle x\rangle+\lambda_f  z}{1 + \lambda_f}   + \frac{1}{\beta} \, \frac{\partial}{\partial x}  \right) p \right]\,\mbox{,}
\end{multline}
where  we use approximation  $\langle x\rangle\approx\int \mbox{d}x \, p  \, x $.  The  stationary solution of  \eqref{eq:fokker-planck}  is $p_s(x) = Z^{-1}\, \exp{\left[- \beta  U  \right]}$, where 
\begin{equation} \label{eq:pot_doublewell}
U(x,  z) = V(x) + \frac{1}{2} \left(x - \frac{\langle x\rangle_s (z)+ \lambda_f \, z}{1+\lambda_f} \right)^2 
\end{equation} 
 is an  effective double well potential  and $Z$ is a normalization constant~\cite{Caruel_2018}.  In \eqref{eq:pot_doublewell} we must use the self-consistence condition  $\langle x \rangle_s = \int \mbox{d}x \, p_s(x) \, x$;  such  closure  can be justified rigorously in the thermodynamic limit~\cite{Dauxois03}. 
 
To develop  the linear response  theory we use as a starting point the  approach  developed in~\cite{Shiino87}.  First, we linearize~\eqref{eq:fokker-planck} to obtain the  propagator 
\begin{multline}\label{eq:fokk_plan_lin_soft}
L \, p  = 
 \frac{\partial}{\partial x}  \left[   \left(  \frac{\partial V}{\partial x} +x  -\frac{\langle x \rangle_s +   z  \, \lambda_f}{1 + \lambda_f}  +  \frac{1}{\beta} \frac{\partial}{\partial x}\right)  p  \right]\,\mbox{.}
\end{multline}
A  perturbation $\delta p(x,t) = p(x,t) - p_{s}(x)$  associated with a small change of the macroscopic strain variable  $\delta z(t)$  will then satisfy a linear equation   
\begin{multline}
\frac{\partial \delta p  }{\partial t} = L  \, \delta p - 
 \frac{1}{1 + \lambda_f} \frac{\partial p_{s}}{\partial x}  \left(\int \mbox{d} x  \, x \,  \delta p     +
 \lambda_f\, \delta z   \right) .
\end{multline}
In particular,  the collective variable   $\langle \delta x \rangle (t)=  \int \mbox{d} x  \, x \, \delta p(x,t)$, will evolve according to 
\begin{equation}
 \label{eq:gen_fluc-diss}
  \langle \delta  x \rangle (t)= 
  \int_{-\infty}^{\infty} \frac{\langle\delta x(t^{\prime})\rangle  + \lambda_f  \delta z(t^{\prime}) }{1 + \lambda_f} \, \chi_{xx}(t - t^{\prime}) \, \mbox{d} t^{\prime} ,
\end{equation}
which in Fourier space reads $\langle \delta  x \rangle (\omega)=\lambda_f \, \chi_{xx}(\omega)\delta z(\omega)/(1+ \lambda_f -\chi_{xx}(\omega)).$  The susceptibility in~\eqref{eq:gen_fluc-diss} is defined by the relation 
\begin{equation}\label{eq:def_chi_xx}
\chi_{x x} (t) = -\Theta(t)\, \int \mbox{d} x \, x \,e^{L  \, t} \frac{\partial}{\partial x} p_{s}(x)\,\mbox{,}
\end{equation}
where $\Theta(t)$ is the Heaviside function.

Next we  write the  conventional fluctuation-dissipation type identity  
$
\chi_{xx} = - \beta \,\Theta(t) \,  dS_{xx}/dt $, where $S_{xx} (t)=  \langle x(t) \, x(0) \rangle   -  \langle x\rangle_s^2$ is  the auto-correlation function  of a single element $x_i$ which is assumed to be evolving in the effective potential~\eqref{eq:pot_doublewell}.  Given that in linear approximation each cross-bridge can be viewed as conducting independently a simple Brownian motion in a double well potential we can use  the  Kramers approximation to obtain~\cite{Skinner78,Pratolongo95}: 
\begin{equation}\label{eq:tau_x1}
S_{xx}(t)  \simeq c  \, e^{-\frac{t}{\tau_{x}}} \, ,    
\end{equation}
where $ 
\tau_{x}\simeq \frac{2 \pi}{\sqrt{|U^{\prime \prime}(x_M)|}}  \frac{e^{\beta U(x_M)}}{\sqrt{U^{\prime \prime}(x_0)} \, e^{\beta U(x_0)}+ \sqrt{U^{\prime \prime}(x_1)} \, e^{\beta U(x_1)}} \, \mbox{,}
$ 
with $x_{0,1}$,  being the minima of the potential~\eqref{eq:pot_doublewell} and $x_M$,  the local maximum between them.  The pre-factor  can be also computed analytically   $c = (x_0-x_1)^2 \, q/(1+q)^2$   with $q= \sqrt{ (U^{\prime \prime} (x_0)/ U^{\prime \prime} (x_1) )} \, \exp[\beta(U(x_0) - U(x_1))] $.  In this computation we neglected the  effect of the relaxation within a single well  because it is  much faster  in physiological conditions than  the barrier crossing.   
\begin{figure}[h]
\includegraphics[width=0.3\textwidth]{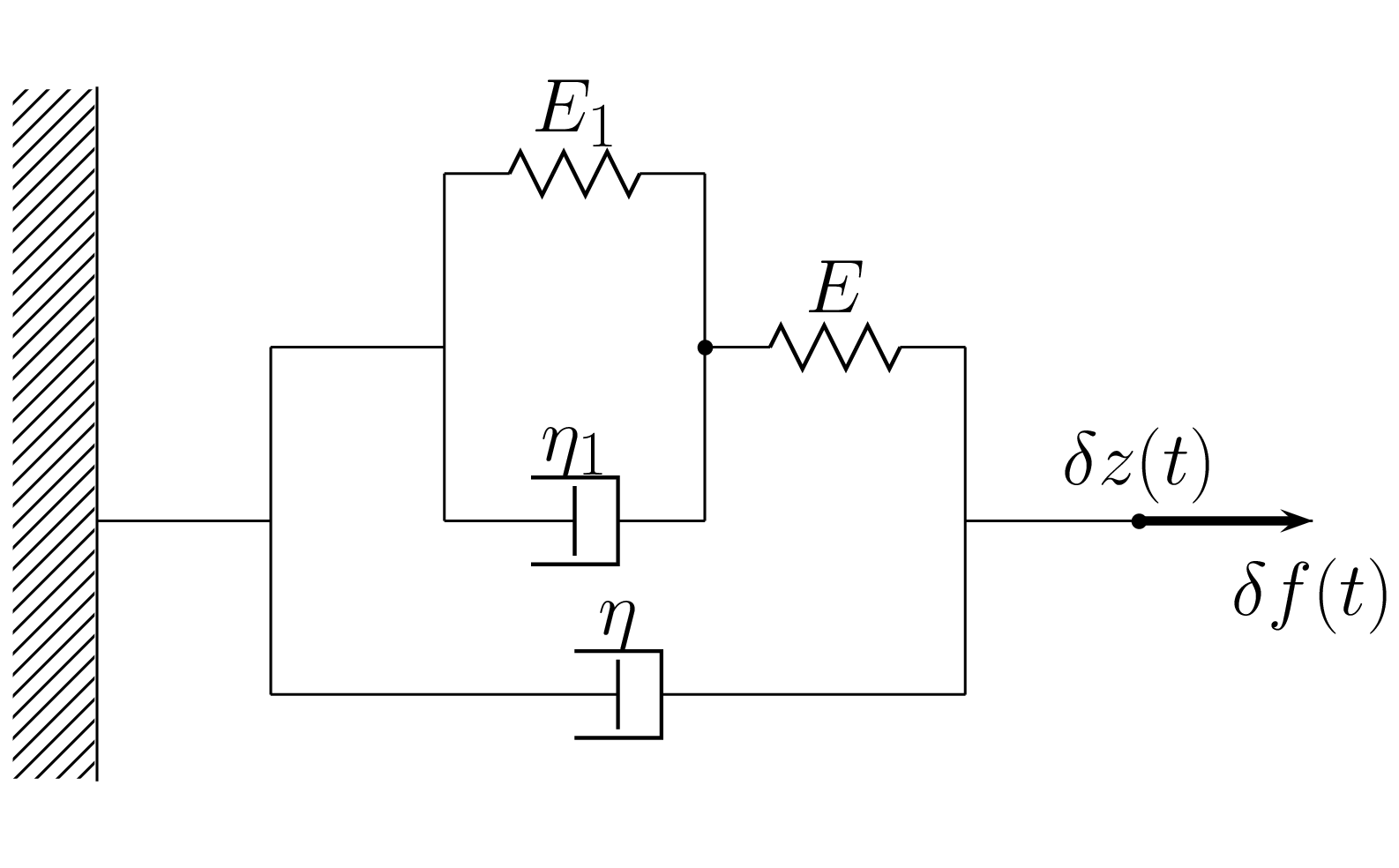}
\caption{Schematic representation of the rheological model~\eqref{eq:constitutive}.} \label{fig:rheomod}
\end{figure}
 
If we rewrite~\eqref{eq:tau_x1} in the Fourier space
$
\chi_{xx} (\omega) =c\,\beta/(1 + i\,\omega \,\tau_x),
$ and  use \eqref{eq:sig_vs_z_dyn}, we  obtain  the desired linear response relation between the macro-variables  $  \delta z(t) $ and $  \delta f (t) $: 
\begin{multline}\label{eq:resp_soft}
\left[ -\omega^2 \tau_x\,\tau_z  +i\,  \omega\left( \tau_z +\tau_x - \frac{\beta\, c\,\tau_z}{1 + \lambda_f } \right) + 1-\beta\,c \right]  \delta z(\omega)  =\\	\left[  i\, \omega\,\tau_x \, \frac{ 1 + \lambda_f}{\lambda_f\,N} + \frac{1+\lambda_f-\beta \, c}{ \lambda_f\,N} \right]  \delta f (\omega)\,\mbox{.}	
 \end{multline}
The corresponding  spring-dash-pot model is shown in Fig.~\ref{fig:rheomod}.   At the inner level we have  a parallel bundle of a spring with stiffness   $E_1 =N\, (1-\beta \,c)/(\beta\,c)$ and a dash-pot with viscosity $\eta_1 = N\,\tau_x/(\beta\,c)$.  This subsystem accounts for  the cross-bridge dynamics.  The outer level   contains a spring with   stiffness $E  =N\, \lambda_f/(1+\lambda_f)$ and the dash-pot with viscosity $\eta  = \nu$; the latter represents  viscous response  of the effective backbone.  

In the real space the rheological  relation \eqref{eq:resp_soft} takes the form
\begin{equation}\label{eq:constitutive}
\theta\,\nu \, \ddot { \delta z}    + (\theta\, E +\nu)\, \dot{  \delta z} + C\, \delta  z = 
    \delta f   + \theta\, \dot{\delta f}  
\end{equation}
where  $C =  E_1 \,E /(E_1 +E )$ and 
$ \theta =  \eta_1 /(E_1 + E )$.
  If $\tau_x=0$ we obtain  the Kelvin-Voigt model, and if $\tau_z=0$  Eq. \ref{eq:constitutive} reproduces the rheological structure of  the (passive) Hill's model.    
  
\begin{figure}[h]
\centering
\label{fig:z_vs_t1}\includegraphics[width=0.5\textwidth]{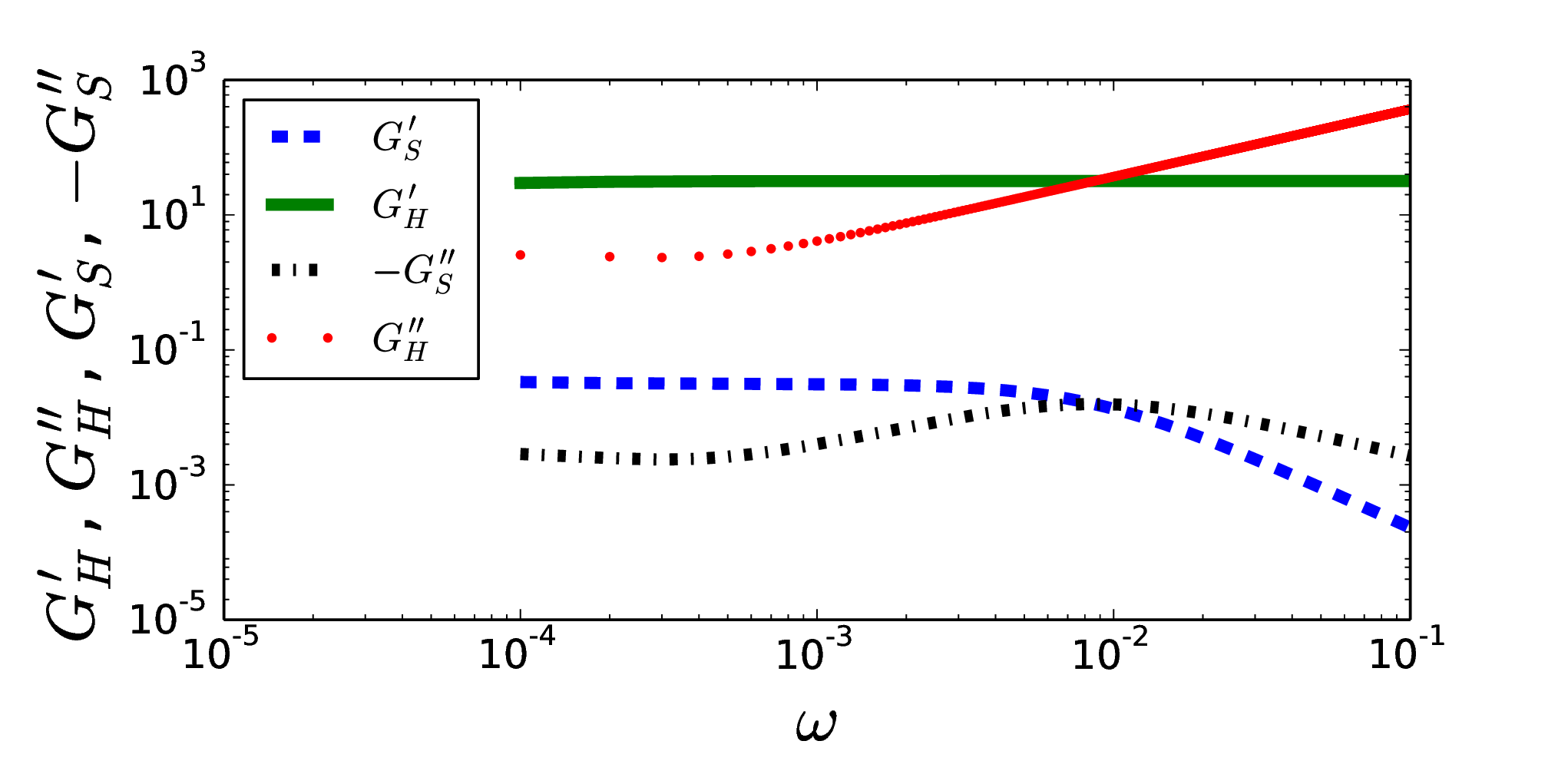}	  
\caption{Frequency dependence of the loss and storage moduli in the  hard and soft devices for the physiological choice of parameters.}
\label{fig:loss_stor}
\end{figure}

To make a more detailed comparison in a hard device we  can  define the storage modulus $G^{\prime}$ and the loss modulus $G^{\prime\prime}$  as  real and  imaginary parts of the ratio $ \delta f(\omega)/\delta z(\omega)$;  in a soft device  we must consider instead  the ratio $ \delta z(\omega)/\delta f(\omega)$. The frequency dependence of these parameters is illustrated in Fig.~\ref{fig:loss_stor}. Note the  divergence of $G^{\prime \prime}_H$ at large $\omega$ in  qualitative difference with  the Hill's model where  this  parameter  tends to zero. Similarly,  in the Hill's model $G^{\prime}_S$ has a finite limit  at large $\omega$ while in our model  it tends to zero.

\begin{figure}[h]
\subfloat[]{\includegraphics[width=0.2 \textwidth]{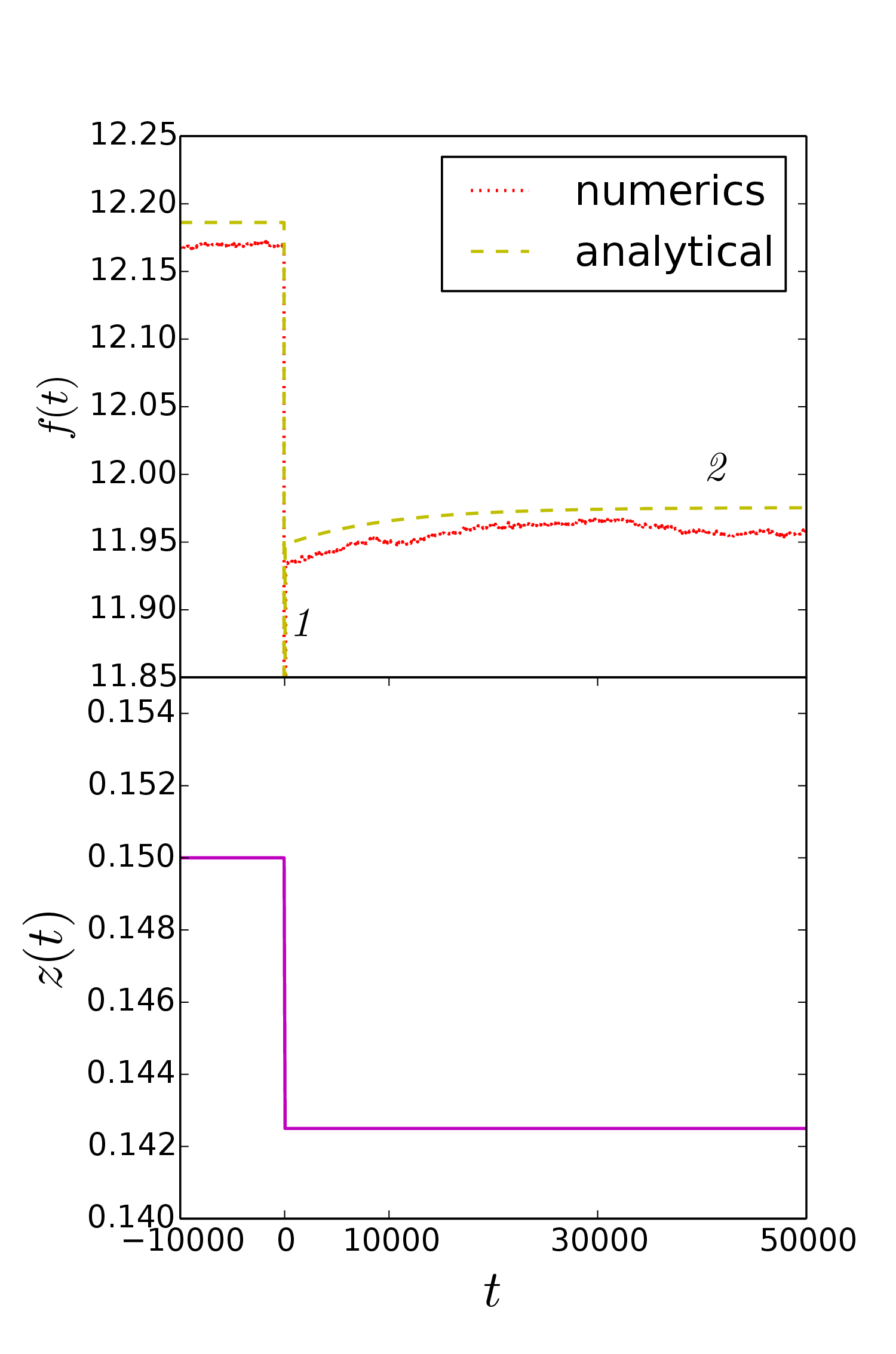}\label{fig:response_hard}}
\subfloat[]{\includegraphics[width=0.2 \textwidth]{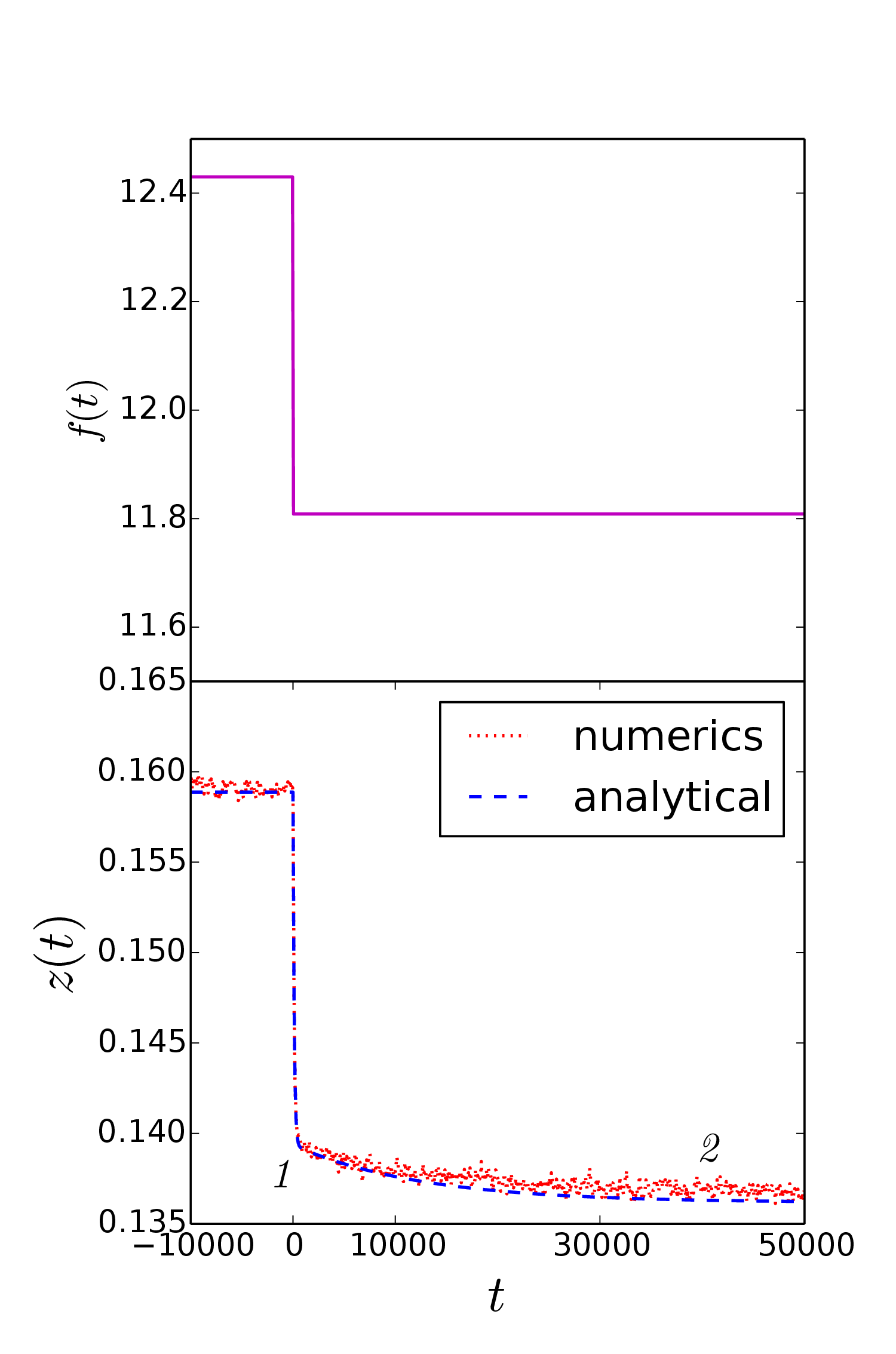}\label{fig:response_soft}}
\caption{Response to a step like perturbation of the model~\eqref{eq:constitutive} (analytical), compared with direct numerical simulations for $N=32768$ crossbridges averaged over 10 realizations (numerics). Here $f$ is the force per thick filament. (a): hard-device, (b): soft device.}
\label{fig:response}
\end{figure}

Consider now  the response of the system~\eqref{eq:constitutive} to  canonical step-like perturbations~\cite{Mainardi2011} as in the typical muscle experiments. In the hard device, the response to the input $ \delta z(t) =   \delta z_0 \, \Theta(t)$, is 
described by the equation
$    \delta f(t)  =   \delta z_0  \left[\, \left(E  -C \right) \, e^{-\frac{t }{\theta }} + C +\nu\, \delta(t) \right]\mbox{,}
$
were $\delta(t)$ is the  Dirac delta function, whose effect cannot be detected in experiments unless the perturbation is strictly instantaneous. Note the first jump in tension $ \lim_{t \rightarrow 0}\delta f(t)   = E \,\delta z_0$, taking place  simultaneously with the applied length step (phase 1 in Fig.~\ref{fig:exp_schem}b and~\ref{fig:response}a), which  is the  signature of a purely elastic behavior~\cite{Reconditi04}. The elastic  phase   is   followed by an exponential relaxation (phase 2 in Fig.~\ref{fig:exp_schem}b and~\ref{fig:response}a)  with the time scale $\theta$.  The condition  $\theta >0$, or equivalently,   $1 + \lambda_f -\beta \,c >0$, then serves as  a bound  (since we are using an approximate expression for the correlation function $S_{xx}$)  for the mechanical  stability threshold of the equilibrium system in the hard device. 

In the soft-device the response to a small step like perturbation $ \delta f(t) =   \delta f_0 \, \Theta(t)$  
is  described by the equation
\begin{multline}\label{eq:resp_step_soft}
   \delta z(t)  =  \delta f_0 \left[\, - \frac{ \, \tau_{+}\, \tau_{-} }{\nu \,(\tau_{+} - \tau_{-}) }\, \left( 1-\frac{\tau_{+}}{\theta}\right) \,e^{-\frac{t}{\tau_{+}}} \right. \\
+\left. \frac{ \tau_{+}\,\tau_{-} }{\nu\,(\tau_{+} - \tau_{-}) }\, \left( 1-\frac{\tau_{-}}{\theta}  \right) \,e^{-\frac{t}{\tau_{-}}} + \frac{1 }{C} \right] \mbox{,}
\end{multline}
where we introduced two new effective time scales
\begin{equation}\label{eq:timescales_soft}
\tau_{\pm} = 
 \frac{\eta_1+\nu}{2\,E_1}+\frac{\nu}{2\,E }\pm \sqrt{\left(\frac{\eta_1+\nu}{2\,E_1}-\frac{\nu}{2\,E } \right)^2 +\frac{\nu^2}{E_1 \,E }} \,\mbox{.}
\end{equation}

Observe that according to~\eqref{eq:resp_step_soft}, $\lim_{t\rightarrow 0}  \delta  z(t)  = 0$, indicating that there is no synchronous  response to a step-like perturbation. The fact that in the soft device the relaxation starts when the perturbation is still being delivered~\cite{Decostre05}, explains  the difficulty in  separating stages 1 and 2 (shown in Fig.~\ref{fig:exp_schem}a and~\ref{fig:response}b) of the transient response in isotonic conditions:  since  $\tau_{-} \ll\tau_{+}$  the  first relaxation process  with the time scale $\tau_{-}$ was  sometimes   interpreted as purely elastic~\cite{Huxley71,Piazzesi03,Decostre05,Piazzesi07}.  The approximate stability condition is now   $1 - \beta\,c >0$.  Note the difference between  the   stability thresholds  in soft and hard device reflecting the ensemble inequivalence in this mean field system~\cite{Barre01,Caruel13}.
  
Using the basic physiological constraints on  the parameters,  $\beta \ge 0$, $\nu \ge 0$, $\lambda_f\ge 0$,   one can show  that  in stable regimes $\tau_+ >\theta$, which is in agreement with the fact that the relaxation in the soft the device is slower than in the hard device. In unstable regimes   the expression under the square root in~\eqref{eq:timescales_soft} may become negative  which would indicate the possibility of  oscillatory relaxation (in the soft device).  Damped oscillations have been indeed observed in some mechanical experiments~\cite{EdmanCurtin,Sugi1981},  however,  they appeared at   larger timescales where the implied mechanical instability might have been  suppressed actively~\cite{Sheshka16}. 
 
To make quantitative predictions and compare our results with physiological measurements we need to calibrate the model using the physical values of  parameters. 

For the cross-bridge stiffness we take the value  $\kappa= 3.29$~pN/nm~\cite{Ford77}, while the combined stiffness of actin and myosin filaments can be estimated at the value  $\kappa_f \simeq 153$~pN/nm~\cite{HUXLEY19942411,WAKABAYASHI19942422}.  Since $N \simeq 100$  we obtain $\lambda_f = \kappa_f/(\kappa \,N)\simeq 0.46$.  Given that the maximum working stroke is $a = 11$~nm, the energy scale is $\kappa \,a^2 = 398  \, \mbox{zJ}$, and we can conclude that at $T = 273$~K the value of the dimensionless inverse temperature is $\beta  \simeq 105 $. 
 
To estimate parameter $\gamma_x$ we assume that a typical  myosin head has the diameter $h \simeq 6$~nm and that the cytoplasm has the effective dynamical viscosity $\eta \simeq 2.3 \times 10^{-3} \,\mbox{Pa}\,\mbox{s}$~\cite{Kushmerick1297,ARRIODUPONT19962327}. Viewing it as a sphere  we obtain $\gamma_x \simeq 3 \, \pi\,\eta \,h \simeq 1,30 \times 10^{-4}$ ms pN/nm, so that the corresponding time unit is $ \gamma_x/\kappa = 3.44 \times 10^{-5}$~ms.  The dimensionless viscosity of a thick filament $\nu= \gamma_z/\gamma_x$ can be now calculated using the estimate for the drag coefficient  $\gamma_z =  \xi/\rho_f \simeq 0.480 \,\mbox{ms pN}/\mbox{nm}$, where $\xi \simeq 3 \times 10^8\, \mbox{N s}/\mbox{m}^3$ is a viscosity coefficient obtained  in~\cite{Ford77} for \textit{Rana temporaria}. We  use here the expression $\rho_f = 2/(\sqrt{3} \,d^2)$ for the density of thick filaments in the section of a sarcomere where  the thick filaments form a triangular pattern and lie at a distance $d=43$~nm~\cite{Matsubara72}.
 The drag coefficient for   the bundle of thick filaments constituting  the half-sarcomere can be taken in the form $\gamma_z^{HS} \simeq \xi \, S$, where $S\simeq 8\, \mu\mbox{m}^2$ is the cross section of the sarcomere. Substituting this expression into the estimate for $\gamma_z$ we can extend the  predictions of the  model to the case of  the whole  half-sarcomere. The number of crossbridges $N$ has to be then multiplied by number of thick filaments in a half-sarcomere with the parallel rescaling $\delta f \rightarrow \delta f/( S\,\rho_f)$ .

To model the double well potential $V(x)$ we use the simplest quartic function with   one minimum $V(0)=0$ (pre-power-stroke state)  and  another minimum   $V(-0.25) =-0.03$ (post-power-stroke state).   
The  barrier  is chosen  at the level  $V(x_*) = 0.11$ because  the  activation energy for the muscle power stroke was previously estimated to be  either  $55.3$~zJ or $95.1$~zJ~\cite{Elangovan12,ANSON92} and we have chosen the smaller value as more relevant  for the  the fast time transient response. 

The rheological equation~\eqref{eq:constitutive} obtained for  a single  half-sarcomere  must be now renormalized  to the scale of a muscle fiber.  To this end we assume that the response is affine, at least  when  perturbations are sufficiently small.  We can view a myofibril  as a  chain of $L\sim 10^4$ half-sarcomeres connected in series, and represent a muscle fiber by a parallel arrangement of $M\sim 200- 2000$ such myofibrils.  The renormalization will then reduce to the substitutions $\delta z \rightarrow  \delta z / L$ and $\delta f \rightarrow \delta f / M$.

Using these values of parameters  we compute $\theta \simeq 10200$ ($\simeq 0.35$~ms) which is compatible with the relaxation time measured in \cite{Piazzesi03}. Next we  estimate the elastic modulus for the   instantaneous response of   a half-sarcomere  in the hard device. Given  that $l_0 =95.5$ ($\simeq 1.05\,\mu\mbox{m}$)  is the length of a half-sarcomere and  $r \simeq 0.83$ is the ratio of the cross-section of the muscle fiber occupied by sarcomeres~\cite{Mobley31}, we obtain $E_Y = \lim_{t \rightarrow 0}  \delta f (t) \,\rho_f\,r\, l_0/\delta z_0\simeq N\,\rho_f\,r\, \lambda_f\,l_0/(1+\lambda_f) = 190$ ($ \simeq 57 \,\mbox{MPa}$), close to the value measured in~\cite{Piazzesi03}. Finally we  compute   $\tau_{+} \simeq 12700$ ($ \simeq 0.44$~ms) and $\tau_{-}\simeq 116$ ($ \simeq 0.004$~ms), which is also in good agreement with experimental observations~\cite{piazzesi2002size,Decostre05}.

We are now in the position to evaluate to what extent the  microscopic stochastic model and the macroscopic deterministic model can  reproduce the outcomes of the  realistic experiments. Numerical simulations of the microscopic model   were conducted with a second order stochastic Runge-Kutta algorithm. We simulated the response of the microscopic model  to a  step-like perturbation   $\delta z$ in a hard device, and $\delta f$ in a soft device, computing the corresponding responses  $\delta f(t)$  and   $\delta z(t)$.  

The  results are summarized in Fig.~\ref{fig:response} where we used  the physiological values of parameters except that in the simulation of the stochastic micro-model of a half sarcomere, instead of the realistic value $N \sim 500000$, we used  the computationally reachable value  $N=32768$,  while  appropriately rescaling the parameter $\nu$ to ensure that the time scale $\tau_z$ remains at its realistic value.
Numerical experiments  aimed at a single bundle of thick and thin filaments with  $N=128$ and using the physiological value of $\nu$ give practically the same results.    In both cases the agreement between the stochastic model and the rheological equation ~\eqref{eq:constitutive} is excellent for both hard and soft devices.

To conclude, we have shown that the time dependent passive viscoelastic behavior of striated  muscles can be understood at the quantitative level starting from the  microscopic structure of a half sarcomere. From a stochastic microscale model we derived a deterministic  rheological  model which describes linear  response of muscle fibers under various loading conditions  and found explicit relations between   the macroscopic and the microscopic parameters. The fact that the derived rheological relation  involves not only first but also second derivatives   allowed us to explain the qualitative differences in  the mechanical response of isometrically and isotonically loaded  muscles.   The model   was calibrated based on independent data and excellent  quantitative agreement was reached with physiological observations. It would be of interest to check experimentally  the  predicted  link between  the  the mechanical response to fast perturbations\cite{piazzesi2002size,Piazzesi03}  and the  power spectrum of mechanical fluctuations \cite{Malmerberg15}.

\emph{Acknowledgements.} The authors thank M. Caruel, R.  Garcia Garcia and S. Ruffo for helpful discussions. FS was supported by a postdoctoral fellowship  from  Ecole Polytechnique. The work of LT  was supported by the grant  ANR-10-IDEX-0001-02 PSL.

\bibliographystyle{apsrev4-1}

\bibliography{passive_response_a}


\end{document}